\documentstyle[prl,aps,epsf]{revtex}
\begin{document}
\draft

\twocolumn[
\hsize\textwidth\columnwidth\hsize\csname@twocolumnfalse\endcsname

\title{Coulomb blockade in superconducting quantum point contacts}  

\author{D.V. Averin} 

\address{Department of Physics and Astronomy, SUNY at Stony Brook, 
Stony Brook NY 11794}

\date{\today}
\maketitle
\begin{abstract}

Amplitude of the Coulomb blockade oscillations is calculated 
for a single-mode Josephson junction with arbitrary electron 
transparency $D$. It is shown that the Coulomb blockade is suppressed 
in ballistic junctions with $D\rightarrow 1$. The suppression 
is described quantitatively as the Landau-Zener transition in 
imaginary time. 

\end{abstract}

]

Coulomb blockade phenomena in mesoscopic conductors have been 
actively studied during the past few years \cite{b1,b2,b3}. They 
arise from the interplay of discreteness, in units of electron 
charge $e$, of electric charge $Q$ of a small conductor, and 
tunneling into the conductor. Coulomb blockade requires for its 
existence ``localization'' of the charge $Q$, the condition that 
implies that the transparency $D$ of the tunnel barriers isolating 
the conductor is small, $D\ll 1$. In ballistic junctions with 
$D\rightarrow 1$ the charge can move freely in and out of the 
conductor and both the charge quantization and associated Coulomb 
blockade are suppressed. Until now, full quantitative understanding 
of such a suppression has been worked out only for nearly-ballistic 
single-mode junctions between two normal conductors \cite{b4,b5}. 
It has been shown that for conductors with the quasi-continuous 
energy spectrum, amplitude of the Coulomb blockade oscillations 
vanishes as the junction reflection coefficient approaches zero: 
$R=1-D\rightarrow 0$. The aim of this work was to study this 
problem for superconducting junctions, where the situation appears 
to be different. Coulomb blockade oscillations arise in this 
case\cite{b12} from the formation of Bloch bands in the 
Josephson potential $U(\varphi)$ periodic in the Josephson 
phase difference $\varphi$. Since the ballistic junctions also have 
periodic Josephson potential, one could expect that Coulomb 
blockade exists even in the ballistic regime. It is shown below that 
this expectation is incorrect and, similarly to the normal case, the 
Coulomb blockade is completely suppressed when $R\rightarrow 0$.  
The suppression can be described quantitatively as the Landau-Zener 
transition in imaginary time.  
 
Coulomb blockade in superconducting junctions can be conveniently 
discussed as the quantum dynamics of the Josephson phase difference 
$\varphi$. The standard Hamiltonian for quantum dynamics of $\varphi$ 
(see e.g., \cite{b1,b7}) consists of the coupling energy 
$H_c(\varphi)$ of the junction electrodes, which in the case 
of low-transparency junctions reduces to a simple Josephson potential 
$U(\varphi)$, and the charging energy $(Q-q)^2/2C$, where $C$ is the 
junction capacitance, $q$ is the charge injected into the junction 
from external circuit, and $Q$ is the charge transferred through the 
junction. Coulomb blockade manifests itself as periodic oscillations 
of the junction characteristics as a function of the charge $q$ with 
the period $2e$. These oscillations can take place either in time 
\cite{b12,b11}, when the junction is biased with a dc current $I$, 
and $q=It$, or as thermodynamic oscillations \cite{b13}, if one of 
the junction electrodes is an isolated island and the charge $q$ 
is induced on the junction capacitance by external gate voltage 
$V_g$ coupled through a gate capacitance $C_g$: $q=C_gV_g$. In both 
situations, oscillation amplitude is the same and can be found from 
the junction free energy $F(q)$. 

For a single-mode junction with large transparency, studied in this 
work, the coupling energy $H_c(\varphi)$ can be represented 
similarly to the normal case \cite{b4} as a sum of the energies 
$H_{L,R}$ of electrons with momenta $\pm k_F$ moving forward and 
backward through the junction, and a potential $V$ responsible for 
scattering between these two directions of propagation. The energy 
of the forward-moving electrons in a superconductor can be written 
in the standard matrix form: 
\begin{equation} 
H_L= \int dx \Psi_L^{\dagger} (x) \left( \begin{array}{cc}
-i\hbar v_F \partial /\partial x & \Delta(x) \\
\Delta^*(x) & iv_F \partial /\partial x \end{array} \right) 
\Psi_L (x) \, ,  
\label{1} \end{equation} 

\vspace*{-4ex}

\[ \Delta(x) = \left\{ \begin{array}{l} \Delta, 
\;\; x<0, \\ \Delta e^{i\varphi} ,\;\; x>0, \end{array} \right. \] 
where $\Psi_L^{\dagger}=(\psi_{L\uparrow}^{\dagger},\psi_
{L\downarrow})$ is the creation operator for quasiparticles 
with momentum $k_F$, and $v_F$ is the Fermi velocity. $H_R$ is given 
by the same expression with $v_F \rightarrow -v_F$.  The pair 
potential $\Delta (x)$ can be written in the step-like form 
(\ref{1}) under the assumption that the characteristic junction 
length $d$ is much smaller than the superconductor coherence length 
$\hbar v_F/\Delta $. 

Below we limit ourselves to the case of adiabatic phase dynamics  
assuming that all energies, including characteristic charging 
energy $E_C=(2e)^2/2C$ and temperature $T$ are much smaller than 
$\Delta$. The adiabatic condition $E_C\, ,T\ll \Delta$ allows us 
to make several simplification in the junction Hamiltonian. First, 
since in this case quasipartilces are not excited in the junction 
electrodes, the charge $Q$ is carried only by Cooper pairs. This 
means that $Q$ can be expressed directly in terms of the 
Josephson phase difference $\varphi$ \cite{1n}: $Q=-2ei\partial 
/\partial \varphi$. Even more importantly, the energy spectrum of 
electrons moving in the contact can be determined in this regime
treating $\varphi$ as stationary. The Hamiltonian $H_L+H_R$ is 
reduced then to a sum of the quasiparticle energies $\varepsilon_k 
(\varphi)$ of the occupied states, and the total junction 
Hamiltonian can be written as 
\begin{equation} 
H=\frac{1}{2C}(\frac{2e}{i} \frac{\partial}{\partial \varphi} -q)^2 
+\sum \varepsilon_k(\varphi) +V \, . 
\label{2} \end{equation}  

The spectrum of eigenenergies $\varepsilon_k(\varphi)$ is found by 
solving the Bogolyubov-de Gennes (BdG) equations with the 
pair potential $\Delta (x)$ (\ref{1}). It consists of the 
continuum of states at energies outside the gap, $|\varepsilon | > 
\Delta$, and two discrete states in the gap:
\begin{equation}
\varepsilon^{\pm} (\varphi) =\mp \Delta \cos \varphi/2\, , \;\;\;\;
\Psi^{\pm} (x) = \sqrt{\xi /2} \left( \begin{array}{c}
1 \\ \mp e^{-i\varphi /2}  \end{array} \right) e^{\pm i k_F x - 
\xi \mid x \mid } \, , 
\label{3} \end{equation} 
where $\xi = (\Delta/\hbar v_F) \sin \varphi/2$. In all these 
expressions $\varphi \in [0,2\pi]$, and they should be continued 
periodically in $\varphi$ beyond this interval. In the regime of 
classical phase dynamics, only the subgap states (\ref{3}) 
contribute to both the dc Josephson current \cite{b9,b10} and the 
ac current at low voltages \cite{b8n}, and one could expect that 
only they are relevant for quantum phase dynamics. In fact, 
as we will see below, the continuous part of the spectrum plays 
important role in determining the effective potential for the 
evolution of $\varphi$. 

The subgap states merge with the continuum when $\varphi = 
0\,\mbox{mod} (2\pi)$. Equation (\ref{3}) shows that as $\varphi$ 
varies from 0 to $2\pi$ the state with momentum $k_F$ moves across 
the energy gap from the lower half of the continuum, $\varepsilon < 
- \Delta$, to the upper half, $\varepsilon > \Delta$, while the $-k_F$ 
state moves in the opposite direction. The states in the continuum 
also shift up or down in a similar fashion. To see this we calculate 
the variation of the density of states $\rho (\varepsilon)$ with 
evolution of $\varphi$ using the Friedel sum rule (see, 
e.g.,\cite{3n,4n}): 
\begin{equation}
\frac{ \partial\rho (\varepsilon)}{\partial \varphi}  =
\frac{i}{2\pi} \frac{ \partial^2 }{\partial \varphi \partial 
\varepsilon } \ln \mbox{det} S(\varepsilon) \, , 
\label{4} \end{equation}  
where $S(\varepsilon)$ is the scattering matrix for scattering 
off the discontinuity of the pair potential $\Delta (x)$ (\ref{1}).  
Straightforward solution of the BdG equations shows that for $+k_F$ 
states 
\begin{equation} 
S(\varepsilon) = \frac{1}{e^{i\varphi}-a^2} \left( 
\begin{array}{cc} |a| (1-e^{i\varphi}) & (1-a^2) \\
(1-a^2)e^{i\varphi} & |a| (1-e^{i\varphi}) \end{array} \right) \, , 
\label{5} \end{equation} 
where $a(\varepsilon)=\mbox{sign}(\varepsilon ) (|\varepsilon |-
(\varepsilon^2-\Delta^2)^{1/2} )/\Delta$ is the amplitude of 
Andreev reflection from a superconductor. Scattering matrix for 
momentum $-k_F$ is obtained from (\ref{5}) by replacing $\varphi$ 
with $-\varphi$. 

From eq.\ (\ref{5}) we get  
\begin{equation}
\frac{i}{2\pi} \int_{0}^{2\pi} d \varphi \frac{ \partial }{\partial 
\varphi } \ln \mbox{det} S(\varepsilon) = \left\{ \begin{array}{ll} 
1\, , & |\varepsilon| >\Delta \, , \\ 0, & |\varepsilon| = \Delta \, .
\end{array} \right.
\label{6} \end{equation} 
Combined with eq.\ (\ref{4}), this equation means that when $\varphi$ 
increases from 0 to $2\pi$, $+k_F$ states in the continuum move up in 
energy, so that precisely one state is removed from the lower half of 
the continuum, $\varepsilon \leq -\Delta$, and is added to the upper 
half, $\varepsilon \geq \Delta $. Together with the shift of the 
subgap states this means that the whole spectrum of $+k_F$ states 
shifts by one state up in energy. The spectrum of $-k_F$ states 
shifts by one state down. The change in energies in this process is 
infinitesimal for all states besides the two states (\ref{3}) which 
move across the gap and change their energy by $2\Delta$. 

\begin{figure}[htb]
\setlength{\unitlength}{1.0in}
\begin{picture}(2.0,2.2) 
\put(.1,.0){\epsfxsize=2.8in\epsfysize=2.3in\epsfbox{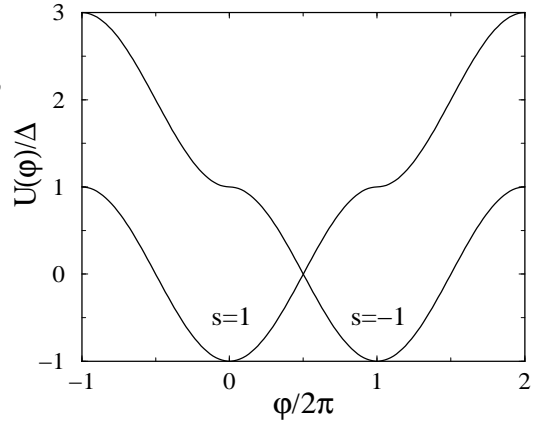}}
\end{picture}
\caption{Two branches of the Josephson potential in a ballistic 
junction with transparency $D=1$: one that corresponds to the 
equilibrium occupation of Andreev states at $\varphi=0$ ($s=1$), 
and another with equilibrium at $\varphi=2\pi$ ($s=-1$). }
\end{figure} 

Such a motion of the energy spectrum determines the effective 
potential for the dynamics of $\varphi$ in the Hamiltonian (\ref{2}). 
At $\varphi =0$, when there are no state in the gap, the equilibrium 
occupation of the eigenstates implies that at $T\ll \Delta$ all the 
states with $\varepsilon \leq -\Delta$ are filled, while those with 
$\varepsilon \geq \Delta$ are empty. Since the adiabatic variation  
of $\varphi$ does not induce transitions between different 
quasiparticle states, the shift of the energy spectrum with these 
occupation probabilities gives rise to the following {\em aperiodic}
potential for $\varphi$ (Fig.\ 1):  
\begin{equation} 
U(\varphi)= \sum \varepsilon_k(\varphi) = \Delta (2m + (-1)^{m+1} 
\cos \varphi/2 )\, ,
\label{7} \end{equation}    

\vspace*{-2ex} 

\[  m\equiv \mbox{int} (\mid \varphi \mid /2\pi)  \, . \]

The rise of the potential (\ref{7}) with $\varphi$ means that the 
phase can increase beyond the points $\varphi =0\,\mbox{mod} 
(2\pi)$ only at the expense of creating quasiparticles in the 
junction electrodes. In the case of classical Josephson dynamics, 
this process generates real quasiparticles and creates dissipative 
component of the Josephson current \cite{b8n}. The energy relaxation 
restores then the $2\pi$-periodicity of all the junction 
characteristics. It should be noted that the potential (\ref{7}) for 
quantum phase dynamics can not be obtained if one takes into account 
only the subgap states \cite{2n}. It is also worth mentioning that 
the mechanism of the spectrum shift creating the potential (\ref{7}) 
is very similar to the mechanism of the chiral anomaly in the 1D 
quantum electrodynamics -- see, e.g., \cite{5n}. 

An important consequence of aperiodicity of the potential 
(\ref{7}) is complete suppression of the Coulomb blockade 
oscillations in ballistic junctions. Since the Coulomb blockade 
in superconducting junctions results from the formation of Bloch 
bands in a periodic Josephson potential, the aperiodicity of the 
potential obviously suppresses the Coulomb blockade. However, the 
periodic nature of the potential and Coulomb blockade are restored 
by finite reflection in the junction. Indeed the aperiodicity 
of the potential (\ref{7}) is the result of the transfer of one 
occupied $+k_F$ states from the energy range $\varepsilon \leq 
- \Delta$ to $\varepsilon \geq \Delta$ and one empty $-k_F$ state 
in the opposite direction as phase evolves from 0 to $2\pi$. The 
backscattering term $V$ in the Hamiltonian (\ref{2}) couples $\pm 
k_F$ states and prevents such a transfer. One can see this by 
looking at the two subgap states (\ref{3}) which in absence of 
backscattering cross the gap range in the course of $\varphi$ 
evolution: the occupied $\psi^+$ state which moves from 
$\varepsilon \leq - \Delta$ to $\varepsilon \geq \Delta$, and the 
empty $\psi^-$ state moving from $\varepsilon \geq \Delta$ 
to $\varepsilon \leq -\Delta$. At $\varphi=\pi$, the energies of 
these states coincide and $V$ couples them effectively. If this 
coupling is sufficiently strong, the occupied state which starts 
off at $\varphi=0$ as $\psi^+$ turns into $\psi^-$ after passing 
the point $\varphi \simeq \pi$ and moves back into the energy range 
$\varepsilon \leq - \Delta$. Similarly, the empty state starting 
at $\varphi=0$ as $\psi^-$ turns into $\psi^+$ and goes back to 
$\varepsilon \geq \Delta$. This means that at $\varphi=2\pi$ all 
states with $\varepsilon \leq - \Delta$ remain occupied while those 
with $\varepsilon \geq \Delta$ remain empty, as at $\varphi=0$. In 
this way the backscattering couples the branch (\ref{7}) of the 
Josephson potential with no quasiparticles at $\varphi=0$ to the 
one with no quasiparticles at $\varphi =2\pi$ (the same as (\ref{7}) 
but shifted along the $\varphi$ axis by $2\pi$ -- see Fig.\ 1), 
thus creating the periodic low-energy branch of the potential. 

Quantitatively, the backscattering term in the Hamiltonian (\ref{2}) 
is $V=\int dx U(x) \rho(x)$, where $U(x)$ is the potential profile 
along the junction and 
\[ \rho (x)=  \sum_{L,R} \Psi_{L,R}^{\dagger}\sigma_3 \Psi_{L,R} +
 (\Psi_R^{\dagger}\sigma_3 \Psi_L e^{2ik_Fx} + h.c. ) \] 
is the operator of electron density. Here and below $\sigma$'s 
denote the Pauli matrices. To evaluate the backscattering term $V$ 
in the basis of the subgap states (\ref{3}) we use the fact that  
the characteristic range of the potential $U(x)$ is defined by the 
junction length $d$ which was assumed to be much smaller than the 
coherence length $\hbar v_F/\Delta$. We find then that the only 
nonvanishing matrix elements are those that couple the two 
branches of the potential: 
\begin{equation}
\langle \Psi^- \!\mid \!V\!\mid \! \Psi^+ \rangle =ir\Delta \sin 
\varphi /2 \, , 
\label{8} \end{equation}
where $r=-iU(2k_F)/\hbar v_F$ is the reflection amplitude of the 
junction \cite{b4}, and $U(2k_F)$ is the Fourier component of 
the potential $U(x)$. At small $r$, the backscattering term (\ref{8}) 
is relevant only in the vicinity of $\varphi=\pi$, where it reduces 
to $ir\Delta$. Then, the junction Hamiltonian (\ref{2}) for $\varphi 
\in [0,2\pi]$ takes the following form in the basis of two branches 
of the potential:  
\begin{equation} 
H=\frac{1}{2C} (\frac{2e}{i} \frac{\partial}{\partial \varphi} -q)^2 
+\Delta (ir\sigma_- -ir^*\sigma_+ -\sigma_3 \cos\varphi/2 ) \, .
\label{9} \end{equation}  

\begin{figure}[htb]
\setlength{\unitlength}{1.0in}
\begin{picture}(2.0,2.1) 
\put(0.0,.0){\epsfxsize=2.8in\epsfysize=2.3in\epsfbox{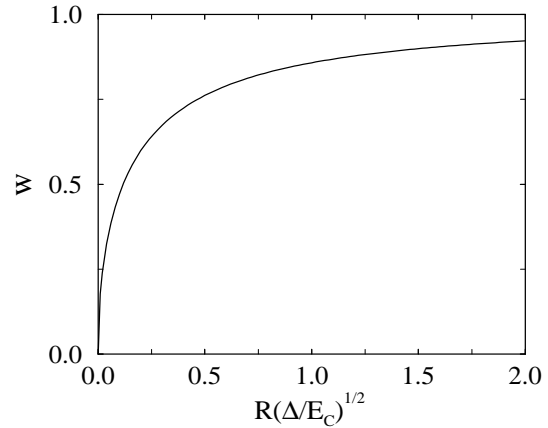}}
\end{picture}
\caption{The probability amplitude $w$ \protect (\ref{11}) for the 
Josephson phase difference $\varphi$ to stay in the low-energy branch 
of the Josephson potential in junctions with small reflection 
coefficient $R\ll 1$.  }
\end{figure} 
The width of the Bloch bands and associated with it amplitude of the 
Coulomb blockade oscillations depend on the probability amplitude 
$w$ of staying on the low-energy periodic branch of the potential 
in the Hamiltonian (\ref{9}). This amplitude is controlled by the 
usual Landau-Zener transition, the same as in the case of classical  
phase dynamics \cite{b8}. The only difference with the classical 
case is that now the transition should take place in the course of 
$\varphi$ motion under the potential barrier, i.e. in ``imaginary 
time''. Indeed, in the quasiclassical approximation, the 
stationary Schr\"{o}dinger equation with the Hamiltonian (\ref{9}) 
and energy $\varepsilon \simeq -\Delta$ describing the evolution of 
$\varphi$ near the level-crossing point $\varphi \simeq \pi$ is:  
\begin{equation}
2 (E_C/\Delta)^{1/2} \partial \psi_s/\partial x = - s x \psi_s/2 
+\sqrt{R} \psi_{-s} \, , 
\label{10} \end{equation}  
where  $x\equiv \varphi-\pi$, and $s=\pm 1$ is the potential branch 
index (Fig.\ 1). In eq.\ (\ref{10}) we removed the phase $\Theta$ 
of the coupling terms in the Hamiltonian (\ref{9}) by the simple 
unitary transformation $\psi_s \rightarrow e^{is\Theta /2} \psi_s$. 
Equations (\ref{10}) are the imaginary-time version of the equations 
describing the regular Landau-Zener transitions, and their solution 
is provided by the parabolic cylinder functions. From the asymptotes 
of these functions \cite{b14} we find that the probability amplitude 
$w$ for the state $s=1$ starting at $x\rightarrow -\infty$ to reach 
the state $s=-1$ at $x\rightarrow \infty$ is: 
\begin{equation} 
w=\frac{1}{\Gamma(\lambda)} \left(\frac{2\pi}{\lambda} \right)^{1/2}
\left( \frac{\lambda}{e} \right)^{\lambda} \, , \;\;\;\; 
\lambda \equiv (R/2)(\Delta/E_C)^{1/2} \, . 
\label{11} \end{equation}  
The amplitude $w$ is plotted in Fig.\ 2. It tends to 
1 at $R\gg (E_C/\Delta)^{1/2}$, while $w\simeq (2\pi \lambda)^{1/2}$ 
at $R\ll (E_C/\Delta)^{1/2}$. 
Since the amplitude of the Coulomb blockade oscillations is 
proportional to $w$, eq.\ (\ref{11}) shows that similarly to the 
normal junctions, in the superconducting case these oscillation 
vanish as $R^{1/2}$ at $R\rightarrow 0$. 

The low-energy periodic branch of the potential in the 
Hamiltonian (\ref{9}) coincides with the classical stationary 
Josephson potential which for arbitrary junction transparency $D$ is 
\cite{b9,b10}: $U(\varphi) = -\Delta [1-D \sin^2 (\varphi/2)]^{1/2}$. 
For $D$ larger than the small ratio $E_C/\Delta$, the charactristic 
magnitude of the potential $U(\varphi)$ is larger than $E_C$, and one 
can find the first few eigenenergies $\varepsilon_n$ for $\varphi$ 
motion in this potential using the quasiclassical wavefunctions away 
from the potential minima at $\varphi=0\, , 2\pi$ and matching them 
to the oscillator wavefunctions in the vicinity of these points. 
Taking into account that the wavefunctions should be 
periodic: $\psi(\varphi+2\pi) = \psi(\varphi)$, we find that each 
oscillator eigenenergy acquires a small correction $-\delta_n$:  
$\varepsilon_n =\hbar \omega_p (n+1/2)-\delta_n$, where 
\begin{equation} 
\delta_0 = \Delta b D w \left(\frac{E_C}{2\pi^2\Delta D}
\right)^{1/4} e^{-a \sqrt{\Delta D/E_C} } \cos \frac{\pi q'}{e} \, , 
\label{12} \end{equation}

\vspace{-2ex} 
\[ \delta_n = (-1)^n \delta_0 \frac{b^{2n}}{n!} 
\left(\frac{\Delta D}{2E_C}\right)^{n/2} \, .   \]
Here $\omega_p = (E_C\Delta D /2\hbar)^{1/2}$ is the frequency of 
small oscillations around the potential minima, and $q'=q- 
e\Theta/\pi$ is the induced charge shifted by the phase of the 
backscattering coupling. The numerical factors $a$ and $b$ in eq.\ 
(\ref{12}) can be expressed in terms of elliptic integrals of the 
first and third kind \cite{b14}, and are plotted as functions of 
the transparency $D$ in Fig.\ 3. At $D\ll 1$, $a=2\sqrt{2}$ and 
$b=4$, while at $D\rightarrow 1$: $a=8(\sqrt{2}-1)+R\ln \sqrt{R}$, 
$b = 8(\sqrt{2}-1)$. The width $\delta_n$ of the Bloch bands  
decreases gradually with increasing $D$ at intermediate $D$'s 
because of the exponential factor in eq.\ (\ref{12}) reflecting 
the increase of the Josephson potential, and then rapidly goes to 
zero at $D\rightarrow 1$ together with the probability amplitude 
$w$ (\ref{11}).    

Summing the corrections $\delta_n$ (\ref{12}) over $n$, we can 
find the $q$-dependent part of the junction free energy at finite 
temperatures ($T\ll \Delta$) on the order of $\hbar \omega_p$: 
\begin{equation} 
F(q) = - \delta_0(q) (1-e^{-\hbar\omega_p/T})  \exp \{-b^2 \left( 
\frac{\Delta D}{2E_C}\right)^{1/2} e^{-\hbar\omega_p/T} \} \, .  
\label{13} \end{equation}  
The free energy (\ref{13}) determines the amplitude of the Coulomb 
blockade oscillations, for instance, oscillations of the voltage 
across the junction: $V(q)=dF(q)/dq$. It should be possible to 
observe the $D$-dependence of the amplitude of the Coulomb 
blockade oscillations (\ref{13}) experimentally either in the 
controllable atomic point contacts \cite{6n} or in the 
semiconductor/superconductor heterostructures \cite{7n}.  

In summary, we have studied the Coulomb blockade oscillations in 
single-mode Josephson junctions with arbitrary electron transparency 
$D$ in the adiabatic limit $E_C\ll \Delta$. It was shown that the 
amplitude of these oscillations decreases steadily with increasing 
$D$ at intermediate $D$'s and then is rapidly suppressed (on the 
scale $(E_C/\Delta)^{1/2}$) at $D\simeq 1$. The rapid suppression 
is described quantitatively by the amplitude (\ref{11}) of the 
Landau-Zener transition between two branches of the Josephson 
potential.   
\begin{figure}[htb]
\setlength{\unitlength}{1.0in}
\begin{picture}(2.0,2.2) 
\put(.1,0.0){\epsfxsize=2.6in\epsfysize=2.3in\epsfbox{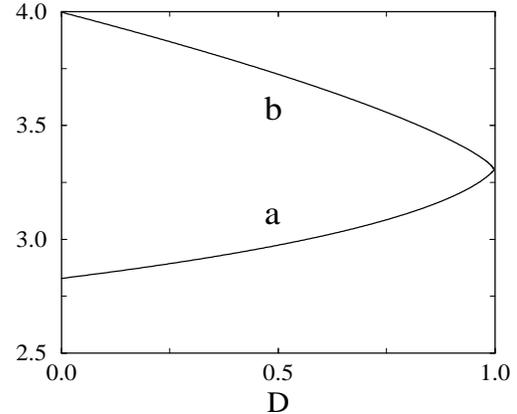}}
\end{picture}
\caption{Exponent $a$ and the preexponential factor $b$ in the amplitude 
of the Coulomb blockade oscillations \protect (\ref{12}) in a single-mode 
Josephson junction as functions of the junction transparency $D$.}  
\end{figure}
The author would like to thank I.L. Aleiner, K.A. Matveev, and 
J. Verbaarschot for useful discussions of the results. This work 
was supported by AFOSR.

\end{document}